\documentstyle[twocolumn,prl,aps,epsfig,floats]{revtex}

\input epsf 
\begin{document}
\twocolumn[\hsize\textwidth\columnwidth\hsize\csname
@twocolumnfalse\endcsname
\input mssymb
\draft
\title{Conformally Invariant Fractals and Potential Theory}
\author{Bertrand Duplantier}
\address{Service de Physique Th\'{e}orique de Saclay, F-91191 Gif-sur-Yvette 
Cedex,\\
 Institut Henri Poincar\'{e}, 11, rue Pierre et Marie Curie, 75231 Paris Cedex
05, France,\\
and Isaac Newton Institute for Mathematical Sciences, 20 Clarkson Road, CB3 OHE, 
U.K.}
\date{19 August 1999}
\maketitle
\begin{abstract}

The multifractal (MF) distribution of the electrostatic potential near any 
 conformally invariant fractal boundary, like a critical 
$O(N)$ loop or a $Q$ -state Potts
cluster, is solved in two dimensions.   
The dimension $\hat f(\theta)$ 
of the boundary set with local wedge angle $\theta$ is 
$\hat f(\theta)=\frac{\pi}{\theta} -\frac{25-c}{12}
\frac{(\pi-\theta)^2}{\theta(2\pi-\theta)}$, with $c$ the central charge of 
the model. 
As a corollary, the dimensions $D_{\rm EP}$ 
of the external 
perimeter and $D_{\rm H}$ of the hull of a Potts cluster obey 
the duality equation $(D_{\rm EP}-1)(D_{\rm  H}-1)=\frac{1}{4}$. 
A related covariant MF spectrum is obtained for self-avoiding walks anchored at 
cluster boundaries.
\end{abstract}
 ]

Fractals are by now a 
well-recognized and studied subject, yet 
they still pose great challenges in mathematical physics: to establish a complete theory of 
diffusion limited aggregates (DLA), or a rigorous theory of the tenuous  
fractal structures arising in critical phenomena, to name only two. 
Classical potential theory, i.e., that of the electrostatic or diffusion
field near such random fractal boundaries, whose self-similarity is reflected in a 
{\it multifractal} (MF) behavior 
of the potential, is almost a {\it Terra Incognita}. In DLA, the potential, 
also called harmonic
measure, actually determines the growth process and its scaling
properties are intimately related to those of the
cluster itself\cite{meak}. In statistical fractals, the Laplacian field is created by the random boundary, and 
should be derivable, in a probabilistic sense, from the knowledge of the latter. A singular example was 
studied in Ref. \cite{cates}, where the fractal boundary, the ``absorber'', was 
chosen to be a simple random walk (RW), or a self-avoiding walk (SAW), accessible to 
renormalization group methods near four dimensions.
 
In {\it two dimensions} (2D), conformal field theory (CFT) has lent strong support to the conjecture  
that statistical systems at their critical point 
produce {\it conformally invariant} (CI) fractal structures, examples of which are 
the continuum scaling limits of RW's, SAW's, critical Ising or Potts clusters, etc. Given the 
beautiful simplicity of the 
classical method of conformal transforms to solve 2D  
electrostatics of {\it Euclidean} domains, perhaps a {\it universal} 
solution is possible for the planar potential near 
a CI fractal.

A first exact example has been 
recently solved in 2D for the whole 
universality class of random or self-avoiding walks, and percolation clusters, which all possess
 the same harmonic MF spectrum \cite{duplantier6} (see also \cite{duplantier4}). 
 In this Letter, I propose 
 the general solution for the potential distribution near any conformal fractal in 2D. 
 The exact multifractal spectra describing the singularities of the potential, or, equivalently, 
 the distribution of wedge angles along the boundary, are obtained, and shown to depend only on 
 the so-called 
 {\it central charge c}, a parameter which labels the universality class of the underlying CFT. 
I devise conformal tools (linked to quantum gravity), which allow the mathematical 
description of random walks interacting with CI fractal structures, 
thereby yielding a complete, albeit probabilistic, 
description of the potential. The results are applied directly to 
well-recognized universal fractals, like $O(N)$ loops or Potts clusters. In particular, a subtle
geometrical structure is observed in Potts clusters, where the {\it external perimeter} (EP), 
which bears the electrostatic charge, differs from the cluster's hull. Its fractal dimension 
$D_{\rm EP}$ is obtained 
exactly, generalizing the recently elucidated case of percolation \cite{DAA}. 

{\it  Harmonic Measure and Potential}. 
Consider a single (conformally invariant) critical random cluster, generically called ${\cal C}$. Let 
$H\left( z\right) $ be the potential at exterior point $z \in {\rm {\bf  C}}$, with 
Dirichlet boundary conditions 
$H\left({w \in \partial \cal C}\right)=0$ on the outer (simply connected) boundary 
$\partial \cal C$ of $\cal C$, and  
$H(w)=1$ on a circle ``at $\infty$'', i.e., of a large radius
scaling like the average size $R$ of $ \cal C$. 
From a well-known theorem due to Kakutani \cite{kakutani}, $H\left( z\right)$ is identical 
to the {\it harmonic measure}, i.e,
the probability that a random walker launched from 
$z$, escapes to $\infty$ without having hit ${\cal C}$.
The multifractal formalism \cite{bb,hent,frisch,halsey1} 
characterizes subsets ${\partial\cal C}_{\alpha }$ of boundary sites
by a H\"{o}lder exponent $\alpha ,$ and a Hausdorff 
dimension $f\left( \alpha \right) ={\rm dim}\left({\partial\cal C}_{\alpha }\right)$, such that their 
potential locally scales as 
\begin{equation}
H\left( z \to w\in {\partial\cal C}_{\alpha }\right) \approx \left( |z-w|/R\right) ^{\alpha },
\label{ha}
\end{equation}
in the scaling limit $a \ll r=|z-w| \ll R,$ with $a$ the underlying lattice constant.
In  2D the {\it complex} potential $\varphi(z)$ (such that the electrostatic potential $H(z)=\Re \varphi(z)$ 
and field $|{\bf E}(z)|=|\varphi'(z)|$) reads for a {\it wedge} of angle $\theta$, centered at $w$:  
$\varphi(z) = (z-w)^{{\pi}/{\theta}}.$ By Eq. (\ref{ha}) a 
H\"older exponent $\alpha$ thus defines a local angle
$\theta={\pi}/{\alpha},$ and the (purely geometrical) MF dimension $\hat f(\theta)$ 
of the boundary subset with such $\theta$ is 
$\hat f(\theta) = f(\alpha={\pi}/{\theta}).$ 

Of special  
interest are the moments of $H$, averaged over all
realizations of ${\cal C}$ 
\begin{equation}
{\cal Z}_{n}=\left\langle \sum\limits_{z\in {\partial {\cal} C(r)}}H^{n}\left( z\right)
\right\rangle ,  
\label{Z}
\end{equation}
 where $\partial {\cal C}(r)$ is shifted a distance $r$ outwards from $\partial \cal C$, and where $n$ can be 
 a real number. In the scaling limit, one expects these moments to scale as 
\begin{equation}
{\cal Z}_{n}\approx \left( r/R\right) ^{\tau \left( n\right) },  \label{Z2}
\end{equation}
where the multifractal scaling exponents 
$\tau \left(
n\right) $ encode generalized dimensions, $D\left( n\right)=\tau \left( 
n\right) /\left( n-1\right)$, 
which vary in a non-linear way with $n$\cite{bb,hent,frisch,halsey1}; 
they obey the symmetric Legendre transform 
$\tau \left( n\right)
+f\left( \alpha \right) =\alpha n,$ with $n=f'\left( \alpha \right), \alpha =\tau'\left( n\right)$.  
From Gauss' theorem \cite{cates} $\tau (1)=0.$ 
Because of the ensemble average (\ref{Z}), values of $%
f\left( \alpha \right) $ can become negative for some domains of $\alpha $ 
\cite{cates}. This Letter is organized as follows: I first present the main 
results in a universal way, 
then proceed with their 
derivation from conformal field theory, and finally specify them for the $O(N)$ and Potts models.

{\it Exact Multifractal Dimensions and Spectra}. 
Each conformally invariant random system
 is labelled by its {\it central charge} $c$, $c\leq 1$. 
 The multifractal dimensions of a simply connected CI boundary then read explicitly:
\begin{eqnarray}
\tau\left( n\right) &=&\frac{1}{2}(n-1)+\frac{25-c}{24}
\left(\sqrt{\frac{24n+1-c}{25-c}}-1\right), 
\nonumber\\
D\left( n\right) &=&\frac{\tau\left( n\right)}{n-1}=\frac{1}{2}+
{\left(\sqrt{\frac{24n+1-c}{25-c}}+1\right)}^{-1}, 
\label{D''}\\
\quad n&\in& \left[ n^{\ast}=
-\frac{1-c}{24}
,+\infty \right) .  
\nonumber
\end{eqnarray}
The Legendre transform reads 
\begin{eqnarray}
\alpha &=&\frac{d{\tau} }{dn}\left( n\right)=\frac{1}{2} +\frac{1}{2}
\sqrt{\frac{25-c}{24n+1-c}};
\label{a'}
\\ 
f\left( \alpha \right)-\alpha&=& \frac{25-c}{24}
\left[1-\frac{1}{2}\left(2\alpha -1 + \frac{1}{2\alpha -1}\right)\right],
\label{f''}
\\  
\quad \alpha &\in& \left( 
{\textstyle{1 \over 2}}%
,+\infty \right) . 
\nonumber
\end{eqnarray}
Notice that the generalized dimensions $D(n)$ satisfy, for any 
$c$, $\tau'(n=1)=D(n=1)=1$, or 
equivalently $f(\alpha=1)=1$, i.e., {\it Makarov's theorem} \cite{mak}, 
valid for any simply connected boundary curve. From (\ref{D''},\ref{a'}) we also remark a fundamental relation, 
independent of $c$:
$3-2D(n)=1/\alpha=\theta/\pi.$
We also have the {\it superuniversal} bounds: $\forall c, \forall n,\frac{1}{2}=D(\infty) \leq D(n) 
\leq D(n^{\ast})=\frac{3}{2}$, 
hence $0 \leq \theta\leq 2\pi$. 
We arrive at the geometrical multifractal distribution 
of wedges $\theta$ along the boundary:
\begin{eqnarray}
\hat f(\theta)=f\left(\frac{\pi}{\theta}\right)=\frac{\pi}{\theta}-\frac{25-c}{12}
 \frac{(\pi-\theta)^2}{\theta (2\pi -\theta)}.
\label{fchap}
\end{eqnarray}
Remarkably enough, the second term also describes the contribution by a wedge to the
density of electromagnetic modes in a cavity \cite{BD}.
 The maximum of $f(\alpha)$ corresponds to $n=0$, and gives the dimension $D_{\rm EP}$ of 
 the support of the measure, i.e., 
the {\it accessible} or {\it external perimeter} as
${\sup}_{\alpha}f(\alpha)=f(\alpha(0))=D(0)$: 
\begin{equation}
D_{\rm EP}=D(0)=\frac{3}{2}-\frac{1}{24}\sqrt{1-c}\left(\sqrt{25-
c}-\sqrt{1-c}\right).
\label{D(c)}
\end{equation}
This corresponds to a {\it typical wedge angle}
$\hat\theta=\theta(0)={\pi}/{\alpha(0)}=\pi(3-2D_{\rm EP}).$ 

The multifractal functions $f\left( \alpha \right)-\alpha
=\hat f(\theta)-\frac{\pi}{\theta}$ 
are {\it invariant} when taken at primed variables  
such that
$2\pi=\theta+{\theta}^{\prime}=\frac{\pi}{\alpha}+\frac{\pi}{{\alpha}^{\prime}}$, 
which corresponds to the complementary domain of the wedge $\theta$. 
This condition reads also $D(n)+D(n')=2.$ 
This basic symmetry, first observed and studied in \cite{BDH} for the $c=0$ result of \cite{duplantier6}, 
is valid 
for {\it any} conformally invariant boundary. 

In Fig. 1 are displayed the multifractal functions $f$, Eq. (\ref{f''}),  
corresponding to various values of 
$-2 \leq c \leq 1$, or, equivalently, to a number of components 
$N \in [0, 2]$, and $Q \in [0,4]$ in the $O(N)$ or Potts models (see below).  
\begin{figure}[t]
\centerline{\epsfig{file=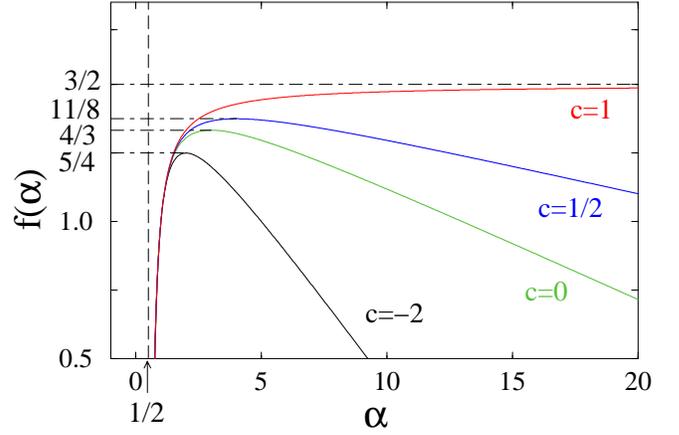,width=8.5cm}}
\caption{Universal harmonic multifractal spectra
$f(\alpha)$. The curves are indexed by the central charge $c$, and correspond
respectively to: 2D spanning trees ($c=-2$); self-avoiding or random walks, and 
percolation ($c=0$); 
Ising clusters or $Q=2$ Potts clusters ($c=\frac{1}{2}$); $N=2$ loops, or $Q=4$ 
Potts clusters 
($c=1$). The maximal dimensions are those of the {\it 
accessible} frontiers. The left branches of the various $f(\alpha)$ curves are largely
indistinguishable, while their right branches split for large $\alpha$, 
corresponding
 to negative values 
of $n$.} 
\label{Figure1}
\end{figure}
The generalized dimensions $D(n)$ (not shown) appear as quite similar for various values of 
$c$, and a numerical simulation would hardly distinguish the different universality 
classes, while the $f(\alpha)$ functions do distinguish these classes, especially for negative $n$. 
The singularity at $\alpha=\frac{1}{2}$, or $\theta=2\pi$, in the multifractal functions $f$, or $\hat f$, 
corresponds to boundary points with a needle local geometry, and Beurling's theorem \cite{ahlfors}
indeed insures the H{\"o}lder 
exponents 
$\alpha$ to be bounded below 
by $\frac{1}{2}$. This corresponds to large values of $n$, where, asymtotically, for {\it any}
 universality class,
$ \forall c, \lim_{n \to \infty} D(n)=\frac{1}{2}.$

The right branch of $f\left( \alpha \right) $ has a linear asymptote 
$\lim_{\alpha \rightarrow \infty} f\left(\alpha \right)/{\alpha} = n^{\ast}=-(1-c)/24.$ 
The {\it limit} multifractal spectrum is obtained for $c=1$, which exhibits an {\it exact} example 
of a {\it left-sided} MF spectrum, with an asymptote 
$f\left(\alpha \to \infty, c=1\right)\to \frac{3}{2}$ (Fig. 1). It corresponds to singular boundaries where 
${\hat f}\left(\theta \to 0, c=1\right)=\frac{3}{2}=D_{\rm EP}$, i.e., where the 
external perimeter is dominated by ``{\it fjords}'', with typical angle $\hat \theta =0$.

The $\alpha \to \infty$ behavior corresponds to moments 
of lowest order $n\rightarrow {n^{\ast}}$, where 
$D(n)$ reaches its maximal value: $\forall c, D(n^{\ast})=\frac{3}{2}$, 
common to {\it all} simply connected, conformally invariant, boundaries. 
This describes almost inaccessible sites: 
define ${\cal N}\left( H\right)$ as the 
number of boundary sites
having a given
probability $H$ to be hit by a RW starting at infinity;
the MF formalism yields,  
for
$H\rightarrow 0,$ a power law behavior
${\cal N}\left( H\right)|_{H\rightarrow 0}\approx H^{-(1+{n}^{\ast})}$
with an exponent $1+n^{\ast}=\frac{23+c}{24}<1.$ 

{\it Conformal Invariance and Quantum Gravity.} Let me
 now give the main lines of the derivation of exponents $\tau\left(
n\right) $, hence $f(\alpha),$ by generalized {\it conformal invariance}. 
By definition of the $H$-measure, $n$ {\it independent} RW's, or Brownian paths ${\cal B}$ 
in the scaling limit, starting at the same 
point a distance $r$ away
from the cluster's hull $\partial \cal C$, and diffusing 
without hitting $\partial \cal C$, give a geometric representation of the $n^{th}$
moment$, H^{n},$ in Eq.(\ref{Z}) for $n$ {\it integer}. Convexity yields analytic
continuation for arbitrary $n$'s. Let us introduce the notation $A\wedge B$ for 
two random sets conditioned to traverse, {\it without mutual intersection}, the 
annulus ${\cal%
D}\left( r, R\right) $ from the inner boundary circle of radius $r$ to the outer 
one at distance $R$,
and{\it \ }$A\vee B$ for two {\it independent}, thus 
possibly
intersecting, sets \cite{duplantier6}. With this notation, the probability that the Brownian paths 
and cluster are in a configuration ${\partial \cal C}\wedge 
\left( \vee {\cal B}\right) ^{n}\equiv {\partial\cal C}\wedge {n}$, is expected to scale for $%
R/r\rightarrow \infty $ as 
\begin{equation}
{\cal P}_{R}\left( {\partial\cal C}\wedge n\right) \approx
\left( r/R\right) ^{x\left( n\right) },  
\label{xp}
\end{equation}
where the scaling exponent $x\left(n\right)$ depends on $n.$ 
In terms of definition (\ref{xp}), the harmonic measure moments (\ref{Z}) simply 
scale as
${\cal Z}_{n}\approx R^2{\cal P}_{R}\left( {\partial \cal C}\wedge n\right)$
\cite{cates,duplantier6},  
 which, combined with Eq. (\ref{Z2}), leads to 
$\tau \left( n\right) =x\left( n\right) -2$.  

To calculate these exponents, I use a fundamental mapping of the conformal field theory in the {\it 
plane} $%
{\rm \bf R}^{2},$ describing a critical statistical system, to the
CFT on a fluctuating abstract random Riemann surface, i.e., in presence of {\it quantum gravity}
\cite{KPZ,DK}. Two universal functions $U,$ and $V,$ depending only on the central charge $c$ of 
the CFT, describe this map: 
\begin{eqnarray}
U\left( x\right) &=&x\frac{x-\gamma}{1-\gamma} , \hskip2mm V\left( x\right) 
=\frac{1}{4}\frac{x^{2}-\gamma^2}{1-\gamma}.  \label{U}
\end{eqnarray}
with $V\left( x\right) \equiv U\left( 
{\textstyle{1 \over 2}}%
\left( x+\gamma%
\right) \right)$\cite{duplantiernew}.
The parameter $\gamma$ is the {\it string susceptibility exponent} 
of the random 2D surface (of genus zero),
bearing the CFT of central charge $c$\cite{KPZ}; $\gamma$ is the solution of
$c=1-6{\gamma}^2(1-\gamma)^{-1}, \gamma \leq 0.$

For two arbitrary\ random sets $A,B,$  
with boundary scaling exponents in the {\it half-plane} $\tilde{x}\left( A\right) ,\tilde{x}\left(
B\right),$ the scaling exponent $x\left( A\wedge
B\right)$, as in (\ref{xp}), has the universal structure \cite{duplantier6,duplantiernew} 
\begin{eqnarray}
x\left( A\wedge B\right) &=&2V\left[ U^{-1}\left( \tilde{x}\left( A\right)
\right) +U^{-1}\left( \tilde{x}\left( B\right) \right) \right],  
\label{x}
\end{eqnarray}
where $U^{-1}\left( x\right) $ is the {\it positive} inverse function of $U$
\begin{equation}
U^{-1}\left( x\right) =\frac{1}{2}\left(\sqrt{4(1-\gamma)x+\gamma^2}+\gamma\right) .  \label{u1}
\end{equation} 
$U^{-1}\left( \tilde{x} \right)$ is, on the random Riemann surface, the boundary
scaling dimension corresponding to $\tilde{x}$ in the half-plane  ${\rm \bf R} \times
{\rm \bf R}^{+}$, and the sum of $U^{-1}$ functions in Eq. (\ref{x})
is a {\it linear} representation of the product of two
``boundary operators'' on the random surface, as the condition $A \wedge B$ for two 
mutually avoiding sets is purely {\it topological} there. The sum is mapped back
by the function $V$ into the scaling dimensions in ${\rm \bf R}^2$\cite{duplantiernew}.
 
For the harmonic exponents $x(n) \equiv x\left({\partial\cal 
C}\wedge n \right)$ in (\ref{xp}), we use (\ref{x}). The {\it external 
boundary} exponent $\tilde{x}\left({\partial \cal C}\right)$  
obeys 
$U^{-1}\left( \tilde{x}\right) =1-\gamma$, which I derive either directly, 
or from Makarov's theorem: $\frac{ dx}{ d n}(n=1)=1$\cite{duplantiernew}. The bunch of $n$ 
independent Brownian paths have simply
$\tilde{x}\left( \left( \vee {\cal B}\right) ^{n}\right)=n,$ 
since $\tilde{x}%
\left( {\cal B}\right)=1$ \cite{duplantier6}. Thus I obtain
\begin{equation}
x\left( n\right) =2V\left(
1-\gamma 
+U^{-1}\left( n\right) \right).  \label{fina}
\end{equation}
This finally gives from (\ref{U})(\ref{u1}) $\tau(n)=x(n)-2$: 
\[
\tau\left( n\right) =\frac{1}{2}(n-1)+\frac{1}{4}\frac{2-\gamma}{1-\gamma}
[\sqrt{4(1-\gamma)n+{\gamma}^2}-(2-\gamma)]
\]
from which Eq. (\ref{D''}) follows, {\bf QED}.

This formalism immediately allows generalizations. For instance, in place 
of $n$ random walks, one can consider a set of $n$ {\it independent 
self-avoiding} walks $\cal P$, which avoid the cluster fractal boundary, except 
for their common anchoring point. The associated multifractal exponents $ 
x\left( {\partial\cal C}\wedge  \left( \vee {\cal P}\right)^{n} \right)$ are 
given by (\ref{fina}), with the argument $n$ in $U^{-1}$ simply 
replaced by  ${\tilde x}\left( \left( \vee {\cal P}\right) ^{n}\right) =n{\tilde x}
\left( {\cal P}\right) =\frac{5}{8}n $ \cite{duplantier6}. These exponents govern 
the universal multifractal behavior of the moments 
of the probability that a SAW escapes from $\cal C$. One then gets a spectrum $\bar f$ such that
${\bar f}\left(\bar\alpha=\tilde{x} \left( {\cal P}\right)\pi/\theta \right)
= f\left(\alpha=\pi/\theta\right)={\hat f}(\theta)$, thus unveiling the {\it same invariant} 
underlying wedge distribution as the harmonic measure, {\bf QED}.

{\it $O(N)$ and Potts Cluster Frontiers}. 
The $O(N)$ model partition function
is that of a gas 
$\cal G$ 
of 
self- and mutually avoiding {\it loops} on a given lattice, e.g., 
the hexagonal 
lattice \cite{nien}: 
${Z}_{O(N)} = \sum_{\cal G }K^{{\cal N}_{B}}N^{{\cal N}_{P}},$ 
with $K$ and $N$ two fugacities, associated respectively with the 
total number of occupied bonds 
${\cal N}_{B}$, and with the total number 
${\cal N}_{P}$ of loops, 
i.e., polygons drawn on the lattice. 
For $N \in 0.[-2,2]$, this model possesses a critical point (CP), $K_c$, 
while the whole {\it ``low-temperature''} (low-$T$) phase, i.e., ${K}_c < K$,
has critical universal properties, where the loops are {\it denser} 
that those at the critical point\cite{nien}. 
   
The partition function of the $Q$-state Potts model on, e.g., 
the square lattice, with a second order critical point for $Q \in [0,4]$, has a 
Fortuin-Kasteleyn representation {\it at} the CP: 
$ Z_{\rm Potts}=\sum_{\cup (\cal C)}Q^{\frac{1}{2}{\cal N}_{P}},$
where the configurations $\cup (\cal C)$ are those of reunions of 
clusters on 
the square lattice, with 
a total number ${\cal N}_{P}$ of polygons encircling all clusters, 
and filling the medial square lattice of the original lattice \cite{nien,denijs}. 
Thus the critical Potts model becomes a {\it dense} loop model, with a loop fugacity 
$N=Q^{\frac{1}{2}}$, while one can show that its {\it tricritical} point with site 
dilution corresponds to the $O(N)$ CP\cite{D6}. 
The $O(N)$ and Potts models thus 
possess the same ``Coulomb gas'' 
representations \cite{nien,denijs,D6}:
$N=\sqrt{Q}=-2 \cos \pi g,$ 
with $g \in [1,\frac{3}{2}]$ for the $O(N)$ CP, and $ g \in [\frac{1}{2},1]$ for the low-$T$ $O(N)$, 
or critical Potts, models;
the coupling constant $g$ of the Coulomb gas  
yields also the central charge:
$c=1-6{(1-g)^2}/{g}.$ 
\noindent The above representation for $N=\sqrt Q \in [0,2]$ gives 
a range of values $- 2 \leq c \leq 1$; our results also apply for $c \in(-\infty, -2]$, 
corresponding, e.g., to the $O\left(N\in [-2,0]\right)$ branch, with a low-$T$ phase for $g \in [0,\frac{1}{2}]$, 
and the CP for $g \in [\frac{3}{2},2].$
 
The fractal dimension $D_{\rm EP}$ of the accessible perimeter, Eq. (\ref{D(c)}), is, 
like $c(g)=c(g^{-1})$, 
a symmetric function
\begin{equation}
D_{\rm EP}=1+ \frac{1}{2}g^{-1}\vartheta(1-g^{-1})+\frac{1}{2}g \vartheta(1-g),
\label{DEP}
\end{equation}
where $\vartheta$ is the Heaviside distribution, thus 
given by two different analytic
expressions on either side of the separatrix $g=1$. 
The dimension of the {\it hull}, i.e., the complete set of outer boundary 
sites of a cluster, has been 
determined for $O(N)$ and Potts clusters \cite{SD}, and reads 
$D_{\rm H}=1+\frac{1}{2}g^{-1},$
for the {\it entire} range of the coupling constant $g \in [\frac{1}{2},2]$. 
Comparing to Eq. (\ref{DEP}), 
we therefore see that the accessible perimeter and hull dimensions {\it coincide} for $g\ge 1$, i.e., 
at the $O(N)$ CP (or for tricritical Potts clusters), whereas they {\it differ}, namely $D_{\rm EP} <
D_H$, for $g < 1$, i.e., in the $O(N)$ low-$T$ phase, 
or for critical Potts clusters. 
This is the generalization to any Potts model of the effect originally found  
in percolation \cite{GA}. This can be directly understood in terms of the 
{\it singly connecting} sites (or bonds) where fjords close in the scaling limit. 
Their dimension is given by 
$D_{\rm SC}=1+\frac{1}{2}g^{-1}-\frac{3}{2}g\cite{SD}.$
Thus, for critical $O(N)$ loops, $g \in (1,2]$ and $D_{\rm SC} < 0,$ so there exist no closing fjords, 
which explains the 
identity:
$D_{\rm EP} = D_{\rm H};$
whereas $D_{\rm SC} > 0, g \in [\frac{1}{2},1)$ for critical Potts clusters, 
or in the $O(N)$ low-$T$ phase, 
where pinching points of positive dimension appear in the scaling limit, 
so that $D_{\rm EP} < D_{\rm H}$ (Table 1). I then find from Eq. (\ref{DEP}), with $g\leq 1$:
\begin{equation}
\left(D_{\rm EP}-1\right) \left( D_{\rm H}-1\right)=\frac{1}{4}.
\label{duali}
\end{equation}
The symmetry point $D_{\rm EP} = 
D_{\rm H}=\frac{3}{2}$ corresponds to $g=1$, $N=2$, or $Q=4$, 
where, as expected, the dimension $D_{\rm SC}=0$ of the pinching points 
vanishes.  

For percolation, described either by $Q=1$, or by the low-$T$ $O(N=1)$ model, with 
$g=\frac{2}{3}$, we recover the result $D_{\rm EP}=\frac{4}{3}$, recently derived in \cite{DAA}. 
For the Ising model, described either by $Q=2, g=\frac{3}{4}$, or by 
the $O(N=1)$ CP, 
$g'=g^{-1}=\frac{4}{3}$, we observe 
that the EP dimension $D_{\rm EP}=\frac{11}{8}$ coincides, as expected, with that of the critical 
$O(N=1)$ 
loops, which in fact appear as EP's. This is a particular case of a further duality relation 
between the critical Potts and CP $O(N)$ models:
 $D_{\rm EP}\left(Q(g)\right)=
D_{\rm H}\left(O\left(N(g')\right)\right),$ for $g'=g^{-1}, g \le 1$.
\begin{table}[t]
\begin{tabular}{| c | c | c | c | c | c |}
$Q$          &     0     &    1     &      2      &      3      &  4          \\
\hline
$c$          &     -2    &    0     &     1/2     &      4/5       &  1        \\
\hline
$D_{\rm EP}$ & ${5}/{4}$ &${4}/{3}$ & ${11}/{8}$  & ${17}/{12}$ & ${3}/{2}$ \\
\hline
$D_{\rm H}$  & $2$       &${7}/{4}$ & ${5}/{3}$   & ${8}/{5}$   & ${3}/{2}$ \\
\hline
$D_{\rm SC}$ & ${5}/{4}$ &${3}/{4}$ &$ {13}/{24}$ & ${7}/{20}$  &   $ 0$        \\
\end{tabular}
\medskip
\caption{Dimensions for the critical $Q$-state Potts model; $Q=0,1,2$ 
correspond respectively
to spanning trees, percolation and Ising clusters.}
\end{table}  

The gracious hospitality of the Institute for Advanced Studies (Princeton), and of
the Isaac Newton Institute (Cambridge), where this work was carried out, 
is gratefully acknowledged, as is a careful reading of the manuscript by T. C. Halsey.


\begin{references}

\bibitem{meak}
P. Meakin, {\it Fractals, Scaling and Growth Far from Equilibrium}, 
Cambridge Nonlinear Sc. Series {\bf 5} (1999); B.B. Mandelbrot and C.J.G. 
Evertsz, Nature {\bf 348}, 143 (1990).

\bibitem{cates}  M.E. Cates and T.A. Witten, Phys. Rev. A {\bf 35}, 1809 (1987).

\bibitem{duplantier6}  B. Duplantier, Phys. Rev. Lett. {\bf 82}, 880, 3940 (1999).

\bibitem{duplantier4}  B. Duplantier, Phys. Rev. Lett. {\bf 81}, 5489 (1998); 
G.F. Lawler and W. Werner (to be published); J. Cardy, J. Phys. A {\bf 32}, L177 (1999).

\bibitem{DAA}  M. Aizenman, B. Duplantier, and A. Aharony, Phys. Rev. Lett. {\bf 83}, 1359 (1999).

\bibitem{kakutani} S. Kakutani, Proc. Imper. Acad. Sci. (Tokyo) {\bf 20}, 706 (1942).

\bibitem{bb}  B.B. Mandelbrot, J. Fluid. Mech. {\bf 62}, 331 (1974).

\bibitem{hent}  H.G.E. Hentschel and I. Procaccia, Physica (Amsterdam) {\bf %
8D}, 835 (1983).

\bibitem{frisch}  U. Frisch and G. Parisi, in Proceedings of the
International School of Physics ``Enrico Fermi'', course LXXXVIII, edited by
M. Ghil (North-Holland, New York, 1985) p. 84.

\bibitem{halsey1}  T.C. Halsey, M.H. Jensen, L.P. Kadanoff, I. Procaccia,
and B.I. Shraiman, Phys. Rev. A {\bf 33}, 1141 (1986).

\bibitem{mak}  N.G. Makarov, Proc. London Math. Soc. {\bf 51}, 369 (1985).

\bibitem{BD} R. Balian and B. Duplantier, Ann. Physics {\bf 112}, 165 (1978), p.183.

\bibitem{BDH} R. C. Ball, B. Duplantier, and T. C. Halsey, unpublished (1999).

\bibitem{ahlfors}
L. V. Ahlfors, {\it Conformal Invariants. Topics in Geometric 
Function Theory}, McGraw-Hill, New York (1973).

\bibitem{KPZ}  V.G. Knizhnik, A.M. Polyakov, and A.B. Zamolodchikov, Mod.
Phys. Lett. A {\bf 3}, 819 (1988).

\bibitem{DK}  B. Duplantier and I.K. Kostov, Nucl. Phys. {\bf B340}, 491 (1990).

\bibitem{duplantiernew}
B. Duplantier, to be published.

\bibitem{nien}
B. Nienhuis, Phys. Rev. Lett. {\bf 49}, 1062 (1982);
 in {\it Phase Transitions and Critical Phenomena}, edited by C.
  Domb and J. L. Lebowitz, (Academic, London, 1987), Vol. 11.
  
\bibitem{denijs}
M. den Nijs, Phys. Rev. B {\bf 27}, 1674 (1983).

\bibitem{D6}  B. Duplantier, J. Stat. Phys. {\bf 49}, 411 (1987); Phys. Rep. {\bf 184}, 229 (1989).

\bibitem{SD}  H. Saleur and B. Duplantier, Phys. Rev. Lett. {\bf 58}, 2325
(1987).

\bibitem{GA}  T. Grossman and A. Aharony, J. Phys. A {\bf 20}, L1193 (1987).

\end{references}
\end{document}